\documentclass[a4paper]{article}
\usepackage{fourier}%
\usepackage{heuristica}
\usepackage{geometry}
\usepackage{multirow}

\usepackage{tabularx}
\usepackage{caption}
\usepackage{verbatim}
\usepackage{INTERSPEECH2021}

\title{Improve GAN-based Neural Vocoder using Pointwise Relativistic Least Square GAN}
\name{Congyi Wang$^*$\thanks{* Equal Contributor}, Yu Chen$^*$, Bin Wang, Yi Shi}
\address{Xmov, China}
\email{\{wangcongyi,yuchen,wangbin,yishi\}@xmov.ai}

\begin{document}

\maketitle
\begin{abstract}
GAN-based neural vocoders, such as Parallel WaveGAN and MelGAN have attracted great interest due to their lightweight and parallel structures, enabling them to generate high fidelity waveform in a real-time manner. In this paper, inspired by Relativistic GAN\cite{jolicoeur2018relativistic}, we introduce a novel variant of the LSGAN framework under the context of waveform synthesis, named Pointwise Relativistic LSGAN (PRLSGAN). In this approach, we take the truism score distribution into consideration and combine the original MSE loss with the proposed pointwise relative discrepancy loss to increase the difficulty of the generator to fool the discriminator, leading to improved generation quality. Moreover, PRLSGAN is a general-purposed framework that can be combined with any GAN-based neural vocoder to enhance its generation quality. Experiments have shown a consistent performance boost based on Parallel WaveGAN and MelGAN, demonstrating the effectiveness and strong generalization ability of our proposed PRLSGAN neural vocoders.  
\end{abstract}

\section{Introduction}
In recent years, deep neural network based vocoder has been developing rapidly. Compared with conventional vocoders(\cite{morise2016world}, \cite{kawahara2006straight}), neural vocoders can significantly enhance the speech synthesis quality of the current text-to-speech (TTS) system. Most early studies of neural vocoders are based on autoregressive(AR) models, such as Wavenet\cite{oord2016wavenet}, WaveRNN\cite{kalchbrenner2018efficient}, SampleRNN\cite{mehri2017samplernn}, FeatherWave\cite{tian2020featherwave} etc. In such models, samples are generated sequentially while RNNs are utilized in modeling the long-term relationship that existed in the natural waveform. Although they produce very high-quality waves, the generation speed is unfavorable because of the sequential structure, limiting their practical usage in real-time TTS systems. \\
\indent To address the efficiency issue, many approaches are proposed to accelerate the inference speed of AR models, Yu et al.\cite{yu2019durian} modify the original WaveRNN\cite{kalchbrenner2018efficient} and divide the full-band audio signal into four subbands and predict the four subband parameters simultaneously, which leads to reduced prediction duration and model parameters. \cite{valin2019lpcnet} proposes another lightweight vocoder based on the WaveRNN framework. It can synthesize high-quality waveform using linear prediction coefficients (LPC).\\
\indent Recently, non-AR models have drawn increasing attention from researchers. Those models are able to generate a waveform in a highly parallelizable manner to take full advantage of modern hardware with inference speed. Among all the methods, knowledge distill technique plays a major role \cite{oord2018parallel},\cite{ping2018clarinet}. Specifically, the 'knowledge' of one AR teacher model is transferred to a small student model based on the inverse autoregressive flows combined with an extra perceptual loss. Although the resulting model can synthesize high-quality waveform with a reasonable speed, it requires a well-trained teacher model as well as complex training strategies. The gigantic size of their model also restricts it from achieving real-time computation. Another notable endeavor in this field is made by flow-based models including waveglow\cite{prenger2019waveglow}, flowavenet\cite{kim2019flowavenet}, Melflow\cite{zeng2020melglow}, waveflow\cite{ping2020waveflow} and FBWAVE\cite{wu2020fbwave} etc. They apply a single log-likelihood loss to train specially designed invertible models. The inference speed is faster than AR models and can be deployed even on mobile CPU after extra efforts in engineering\cite{wu2020fbwave}. However, its unstable training process and unsatisfying synthesis quality prevent it from being deployed in industrial applications. \\
\indent Some recent works have utilized generative adversarial network(GAN) to train the vocoders. Concretely, the training process of such a model can be summarized as an adversarial game, the generator tries to synthesize the waveform to fool the discriminator, whereas the discriminator distinguishes the difference between the synthesized wave and the ground truth wave. As it reaches a Nash equalization point, the generator is expected to synthesize a high-quality waveform. GAN-based methods(\cite{kumar2019melgan},\cite{yamamoto2020parallel},\cite{yang2020multi},\cite{yang2020vocgan},\cite{tian2020tfgan},\cite{yamamoto2020parallelv2},\cite{zeng2021lvcnet}) are promising given some models can even synthesize waves in real-time on a single GPU and achieve a higher MOS at the same time, suited for actual industrial deployment. In particular, MelGAN\cite{kumar2019melgan} and ParallelWaveGAN\cite{yamamoto2020parallel}(short for PWGAN) are two fundamental GAN-based neural vocoder architectures. Parallel WaveGAN and MelGAN both use auxiliary loss, i.e., multi-resolution STFT loss and feature matching loss, respectively, so they converge signiﬁcantly faster than the original MelGAN\cite{kumar2019melgan}.  MB-MelGAN\cite{yang2020multi} adopts the same idea as Multi-band WaveRNN and synthesizes the subband signals to accelerate the inference speed. VocGAN\cite{yang2020vocgan} modifies the original MelGAN model by appending a hierarchical conditional discriminator to the multi-scale waveform generated in the intermediate layers as well as deepen the receptive field. TFGAN \cite{tian2020tfgan} considers the time-domain loss for generator and discriminator, which encourages the generator and discriminator to learn waveform both in time and frequency domain aiming at eliminating the artifacts in the high frequency, such as metallic sense and reverb in hearing sense. \cite{yamamoto2020parallelv2} proposes a two-way discriminator for voiced and unvoiced parts of the synthesized waveform respectively.\\
\indent Despite the mentioned advancements, the quality of synthesized speech is far from satisfactory. The synthesized audio is prone to have artifacts in the high-frequency domain, while the frequent occurrence of phase mismatch is another challenge hard to neglect. All previous techniques make modifications based on LSGAN\cite{mao2017least}, where the loss of the discriminator is computed using MSE, which ignores the actual score distribution of each wave segment. As illustrated in figure \ref{fig:ra_illustrate}, there exist many truism score distributions that lead to the MSE equilibrium state, most of which will obtain large scores in some positions, leading to high quality, while those with small score will lead to a local audible artifact, affecting the final MOS score. The resulting adversarial gradients from the discriminator may be strongly dominated by the score indicating the global equilibrium state, thus paying less attention to those local possible artifacts. In this paper, inspired by RaGAN\cite{jolicoeur2018relativistic}, we propose PRLSGAN, an enhanced GAN-based architecture for synthesizing waveform, which is supervised by score distribution instead of a single score, leading to the stricter discriminator and thus push the generator to synthesize higher quality wave. The key novelty of our approach is the combination of pointwise relative score discrepancy loss with the conventional MSE loss. The proposed PRLSGAN can be integrated with almost any existing GAN-based neural vocoders, further improving their synthesis quality. In our experiments, we have demonstrated that when combined MelGAN or PWGAN with the PRLSGAN, the resulting model consistently achieves better performance in objective scores such as PESQ, STOI, and subjective MOS score.

\section{Proposed Method}

 
\begin{table*}[!htbp]
\centering%
\begin{tabular}{c|ccccccc}
  \toprule
  \multirow{2}{*}{Vocoder} & Total & Total & \multirow{2}{*}{Length} & \multirow{2}{*}{Optimizer for G} & \multirow{2}{*}{Optimizer for D} & Start\\
   & iterations & batchsize &  &  & & training D \\
  \midrule
  \midrule
  \multirow{3}{*}{Basic PWGAN} & \multirow{3}{*}{500K} & \multirow{3}{*}{16*4} & \multirow{3}{*}{20480} & Radam, lr=1e-4, & Radam, lr=1e-4, & \multirow{3}{*}{100K} \\
   & & & & betas=(0.9, 0.999), & betas=(0.9, 0.999), & \\
   & & & & grad clip=10 & grad clip=1 & \\
  \midrule
  \multirow{3}{*}{Basic MelGAN} & \multirow{3}{*}{220K} & \multirow{3}{*}{64*4} & \multirow{3}{*}{20480} & Adam, lr=1e-3, & Adam, lr=1e-3, & \multirow{3}{*}{50K} \\
   & & & & betas=(0.9, 0.999), & betas=(0.9, 0.999), & \\
   & & & & no grad clip & grad clip=1 & \\
  \bottomrule
\end{tabular}
\caption{Parameters for our basic PWGAN and MelGAN}
\end{table*}   

\subsection{Basic Methods}
\subsubsection{Parallel WaveGAN}
Parallel WaveGAN (PWGAN) \cite{yamamoto2020parallel} can produce high-fidelity waveforms in real-time on a modern GPU. As a GAN-based model, PWGAN consists of a generator (G), a discriminator (D) and multi-resolution STFT auxiliary loss. The generator is a WaveNet-like architecture conditioned on auxiliary acoustic features (e.g., Mel-spectrogram) which transforms the standard Gaussian noise sequence into the high-fidelity waveform in parallel. In PWGAN, a least-squares GAN is adopted to minimize the adversarial ($L_{adv}^{PWGAN}$) as follows:
\begin{equation}
\label{eq:adv}
  L_{adv}^{PWGAN}(G, D) = \mathbb{E}_{z \sim \mathcal{N}(0,I)}[(1-D(G(z)))^2],
\end{equation}

\noindent where z represents the input noise.
Moreover, the discriminator is trained to minimize the adversarial loss ($L_{D}^{PWGAN}$) formulated as:
\begin{equation}
\label{eq:Dis}
  L_{D}^{PWGAN}(G, D) = \mathbb{E}_{x \sim p_{data}}[(1-D(x))^2] + \mathbb{E}_{z \sim \mathcal{N}(0,I)}[D(G(z))^2]
\end{equation}
\noindent where $x$ represents the raw waveform and $p_{data}$ represents the data distribution of the natural samples.
\subsubsection{Multi-resolution STFT loss}
It's hard to build a robust PWGAN while trained with only adversarial losses. In PWGAN, a multi-resolution STFT loss ($L_{sp}$) is adopted to improve the stability of the GAN training. Besides, it also can accelerate the convergence of the training process. STFT loss is the sum of spectral convergence loss ($L_{sc}$) and log STFT magnitude ($L_{mag}$), which are defined as follows:
\begin{equation}
\label{eq:sc0}
  L_{sc}(x, \hat{x}) = \frac{\| \vert STFT(x) \vert- \vert STFT(\hat{x}) \vert \|_{F}} {\| \vert STFT(x) \vert \|_{F}}
\end{equation}
\begin{equation}
\label{eq:sc1}
  L_{mag}(x, \hat{x}) = \frac{1}{N} \| \log \vert STFT(x) \vert- \log \vert STFT(\hat{x}) \vert \|_{1}
\end{equation}
\noindent where $\hat{x}$ represents the generated sample (i.e., G(z)), and $\|.\|_{F}$ and $\|.\|_{1}$ represent the Frobenius norm and the L1 norm, respectively; $|STFT|$ is the stft magnitudes, and N is the number of magnitude elements. 
The multi-resolution STFT loss is the sum of multiple STFT losses with different parameters (i.e. FFT size, window length, and frame shift.), which is defined as follows:\\
\begin{equation}
\label{eq:sc2}
  L_{stft}(x, \hat{x}) = \frac{1}{M} \sum_{m=1}^{M} (L_{sc}^{m} (x, \hat{x}) + L_{mag}^{m} (x, \hat{x})),
\end{equation}
where $M$ is the number of STFT loss. To balance the two loss
terms, we added a hyperparameter $\lambda_{adv}^{PWGAN}$.
The final training loss of PWGAN generator ($L_{G}^{PWGAN}$) is formulated as:
\begin{equation}
\label{eq:sc3}
  L_{G}^{PWGAN}(G, D) = L_{stft}(x, G(z)) + \lambda_{adv}^{PWGAN} L_{adv}^{PWGAN}(G, D),
\end{equation}

\subsubsection{MelGAN}
Basic MelGAN adopts a stack of transposed convolutional blocks to upsample the Mel-spectrogram to match the length of a waveform. To enhance the naturalness, MelGAN uses multiple-scale discriminators that can handle audio at different levels. Basic MelGAN conducts adversarial training with objectives as:
\begin{equation}
\begin{split}
\label{eq:Dis1}
  L_{D}^{MelGAN}(G, D) = \sum_{k=1}^{K} (\mathbb{E}_{x \sim p_{data}}[(1-D_{k}(x))^2] + \\
  \mathbb{E}_{z \sim \mathcal{N}(0,I)}[D_{k}(G(z))^2]),
\end{split}
\end{equation}

\begin{equation}
\label{eq:adv1}
  L_{adv}^{MelGAN}(G, D) = \sum_{k=1}^{K}\mathbb{E}_{z \sim \mathcal{N}(0,I)}[(1-D_{k}(G(z)))^2],
\end{equation}

where x denotes the natural samples, c denotes the acoustic features (e.g., Mel-spectrogram) and z denotes Gaussian noise vector. \\
\indent In addition to the discriminator’s signal, the feature matching objective is also used to train the generator. This objective minimizes the L1 distance between the discriminator feature maps of real and synthesized audio. As suggested in \cite{yang2020multi}, we replace feature matching with multi-resolution STFT loss. Therefore, the final generator loss for MelGAN can be formulated as:
\begin{equation}
\label{eq:sc4}
  L_{G}^{MelGAN}(G, D) = L_{stft}(x, G(z)) + \lambda_{adv}^{MelGAN} L_{adv}^{MelGAN}(G, D)
\end{equation}

\begin{table*}[!htbp]

\centering%
\renewcommand\tabularxcolumn[1]{m{#1}}
\begin{tabular}{c|cccc|c}
\toprule
Method & MCD & FFE & PESQ(wb) & PESQ(nb) & MOS \\
\midrule
Basic PWGAN & 3.417 & 0.040 & 3.181 & 3.505 & 4.026$\pm$0.073 \\
+PRLSGAN & \textbf{3.365} & \textbf{0.038} & \textbf{3.238} & \textbf{3.597} & 4.182$\pm$0.049 \\
\midrule
Basic MelGAN & 3.334 & 0.039 & 3.278 & 3.603 & 4.176 $\pm$ 0.045 \\
+PRLSGAN & \textbf{3.279} & \textbf{0.032} & \textbf{3.351} & \textbf{3.631} & 4.361$\pm$0.039 \\
\midrule
Ground True & 0.0 & 0.0 & 4.5 & 4.5 & 4.687$\pm$0.046 \\
\bottomrule
\end{tabular}
\centering
\captionsetup{justification=centering,margin=2cm}
\caption{The results of ablation study. MCD(dB) and FFE(Hz): the lower, the better. PESQ and MOS(with 95\% confidence intervals): the higher, the better}
\label{tab:results}
\end{table*}

\subsection{Pointwise Relativistic Least Square GAN}
\begin{figure}[t]
  \centering
  \includegraphics[width=\linewidth]{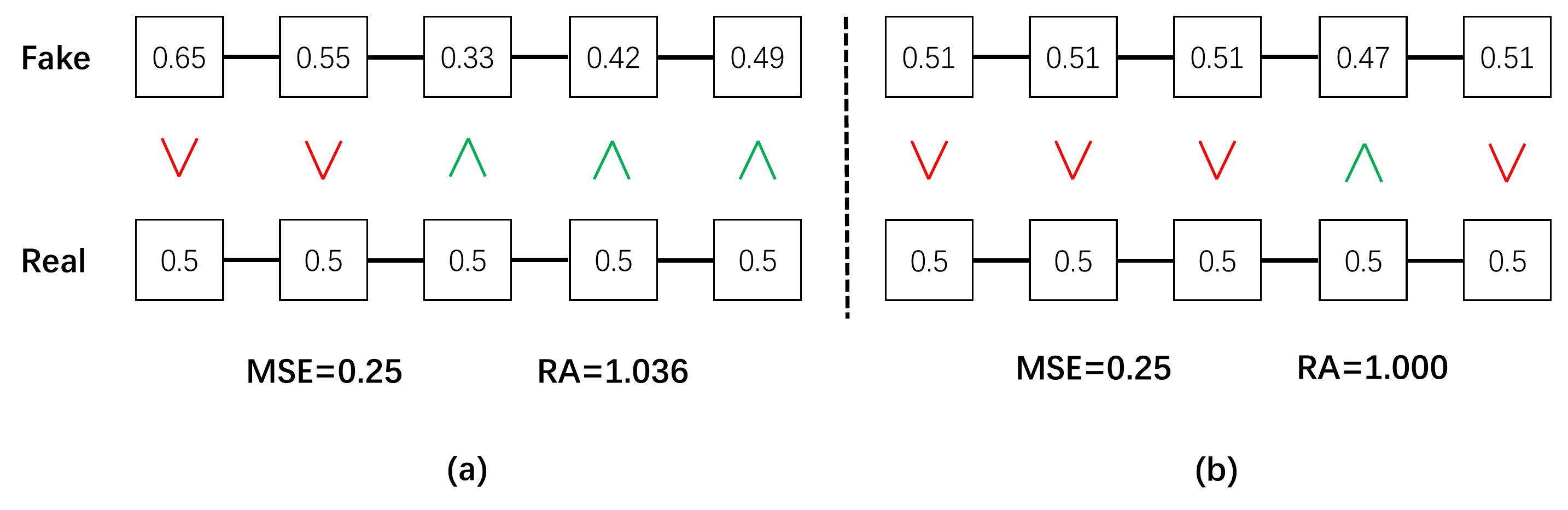}
  \caption{A typical example to illustrate our proposed PRLSGAN. The number in the box is the truism score for each waveform segment, and we compute the MSE and relative pointwise loss(denoted as RA in the figure) for (a) and (b). Although (a) and (b) have the same MSE loss, their relative pointwise loss is different, the score distribution of (b) indicates a higher quality waveform due to its \textbf{relative advantage} over the real data.}
  \label{fig:ra_illustrate}
\end{figure}
\indent The human sense of hearing can be abstracted as a variety of convolution filters in the discriminator in the GAN-based neural vocoders, which are used to judge the wave quality. Our sense of hearing is sensitive to any local audible artifacts when it is surrounded by those seemingly perfect wave segments. However, the least square loss used in the discriminator of LSGAN only considers the average truism scores, making it more likely to ignore the occurrence of local artifacts in its generated wave segments, as illustrated in figure \ref{fig:ra_illustrate}. Note that the synthetic and real waveform in the discriminator input has the same sample length and same semantic meaning. Therefore, we argue that it is vital to consider the score distribution of each wave segment, forcing the generator to avoid the generation of possible local artifacts. In our proposed model, we combine the original MSE loss with the pointwise relativistic discrepancy loss as follows: \\
\begin{equation}
\begin{split}
\label{eq:prlsgan-D}
  L_{D}(G,D)=\mathbb{E}_{x \sim p_{data},z \sim \mathcal{N}(0,I)}[(1-D(x))^2 + D(G(z))^2 + \\ \lambda_{rls}(D(x)-D(G(z))-m)^2]
\end{split}
\end{equation}
\begin{equation}
\begin{split}
\label{eq:prlsgan-G}
  L_{adv}(G,D)=\mathbb{E}_{x \sim p_{data},z \sim \mathcal{N}(0,I)}[\lambda_{adv}(1-D(G(z)))^2 + \\ \lambda_{rls}(D(G(z))-D(x)-m)^2]
\end{split}
\end{equation}

where $\lambda_{rls}$ is experimentally chosen as 0.4, the margin $m$ is set to 1 and $\lambda_{adv}$ is set to 4.0.\\
\indent To further boost the generated waveform quality, we expect the model to refine on obvious artifacts rather than fixing trivial waveform difference which is too subtle to be recognized by human. Specifically, we add a $topK$($K$ is chosen as 10 percent of the whole segment length) loss which emphasizes those large discrepancy wave segments: \\
\begin{equation}
\label{eq:prlsgan-topk-D}
L_{topK}^{D}(G,D)=\frac{1}{K}\sum_{i=1}^{K}[(D(x)-D(G(z))-m)^2]_{(i)}
\end{equation}
\begin{equation}
\label{eq:prlsgan-topk-G}
L_{topK}^{adv}(G,D)=\frac{1}{K}\sum_{i=1}^{K}[(D(G(z))-D(x)-m)^2]_{(i)}
\end{equation}

\noindent where $[\cdot]_{i}$ denotes the $i$th largest number in the scalar sequence.\\
\indent Above all, The final adversarial loss of our PRLSGAN is as follows: \\
\begin{equation}
\begin{split}
\label{eq:prlsgan-final-D}
  L_{D}(G,D)=\mathbb{E}_{x \sim p_{data},z \sim \mathcal{N}(0,I)}[(1-D(x))^2 + D(G(z))^2 + \\ \lambda_{rls}(D(x)-D(G(z))-m)^2 + \lambda_{topK}L_{topK}^{D}(G,D)]
\end{split}
\end{equation}

\begin{equation}
\begin{split}
\label{eq:prlsgan-final-G}
  L_{adv}(G,D)=\mathbb{E}_{(x,c) \sim p_{data},z \sim \mathcal{N}(0,I)}[\lambda_{adv}(1-D(G(c,z)))^2 + \\ \lambda_{rls}(D(G(c,z))-D(x)-m)^2 + 
  \lambda_{topK}L_{topK}^{adv}(G,D)]
\end{split}
\end{equation}

\begin{figure}[t]
  \centering
  \includegraphics[width=\linewidth]{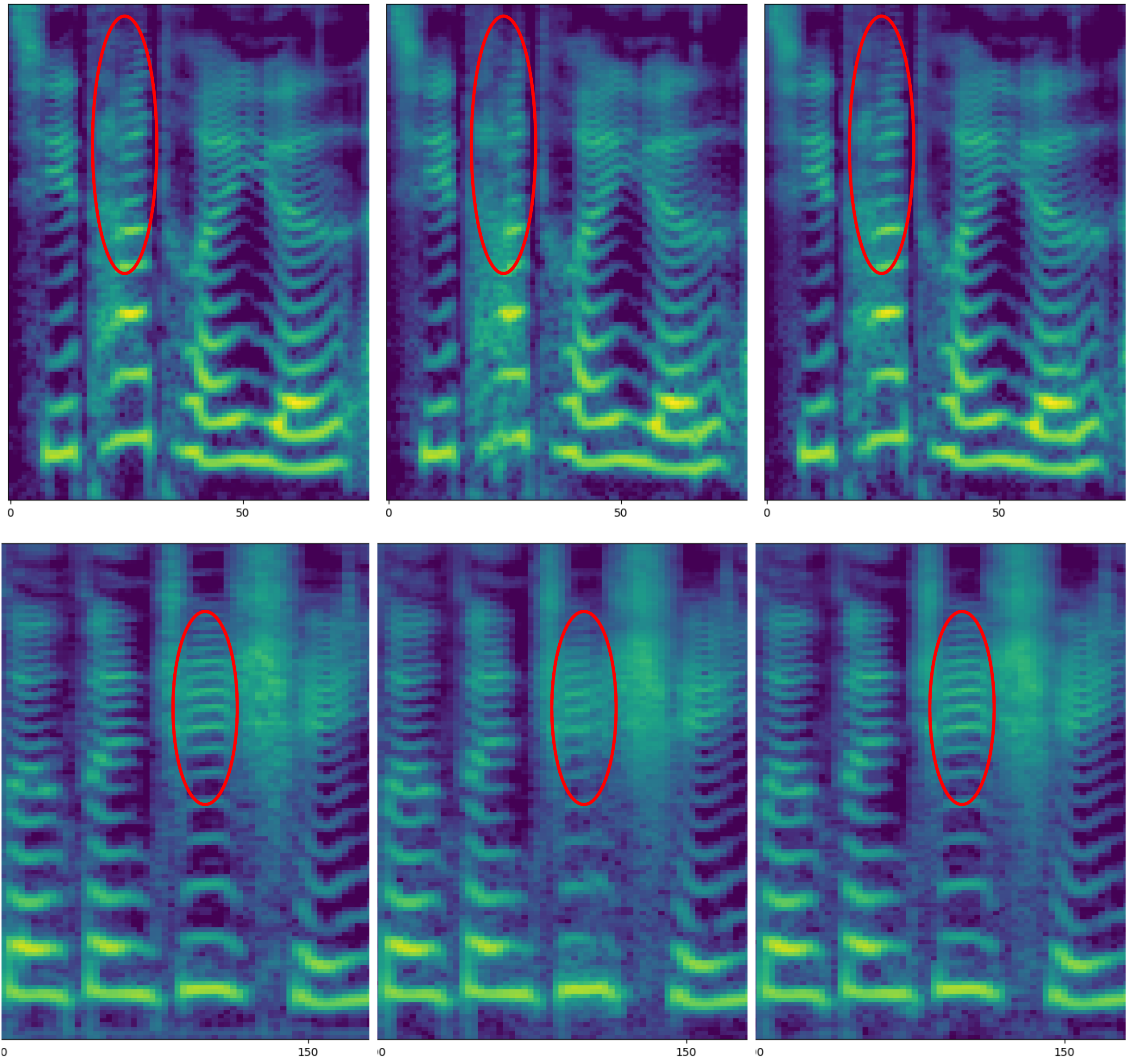}
  \caption{The Comparison between the output Mel-spectrograms of the proposed PRLSGAN and MelGAN based neural vocoders. From left to right column, the ground truth, MelGAN and PRLSGAN.}
  \label{fig:compare}
\end{figure}

\noindent where the $\lambda_{topK}$ is set to 0.01.\\
\indent For the PWGAN, the original MSE loss can be directly replaced by our proposed loss, while for the MelGAN, the same loss must be applied in multiple scales. As demonstrated in table \ref{tab:results}, it is obvious that both the subjective and objective scores are improved with the proposed PRLSGAN framework, comparing with the original LSGAN framework, indicating the effectiveness of our method when implemented in main stream GAN-based neural vocoders. It is also worth noticing in figure \ref{fig:compare}, the high-frequency details are greatly improved by the proposed PRLSGAN framework. 

\section{Experiments}

\subsection{Experiments setup and dataset detail}
We perform experiments on an open-source studio-quality Chinese dataset \footnote{available at www.data-baker.com/open-source.html}, which contains 10,000 audio samples from a Chinese female speaker. The total length of the audio samples are 12 hours. All the recordings were down-sampled to 22050 Hz sampling rate with the 16-bit format. 100 utterances are selected as the test set while the rest samples are used for training. We extracted Mel-spectrograms using 1024-point Fourier transform with 256 hop lengths. 

\subsection{Implementation details}
For the MelGAN structure, we adopt a model called full band MelGAN as proposed in \cite{yang2020multi} which expanded the receptive field by increasing the number of residual blocks to transposed convolutional block. The generator of full-band MelGAN is consists of three transposed convolutional blocks which each block adopted 8, 8, 4 stride, respectively. Each transposed block contains 4 residual blocks with dilated convolutions, and their dilation factors are 1, 3, 9, 27. We did not change the architecture of PWGAN’s generator, and all discriminators.\\
\indent Adam\cite{kingma2014adam} is chosen as the optimizer for MelGAN and RAdam\cite{liu2020variance} for PWGAN respectively. As for the learning rate configuration, for MelGAN, the learning rate of a generator was initialized to 1e-3, reducing by half at 50K, 10K iterations. The discriminator was set to 1e-3 and halved at 10K iterations. For PWGAN, we set the learning rate of the discriminator to 1e-4. For multi-resolution STFT loss, we applied three STFT losses with frame sizes of 512, 1,024, and 2,048, window sizes of 240, 600, and 1,200, and frameshifts of 50, 120, and 240, respectively. Most of the training parameters of our basic models were listed in Table 1. While training models with PRLSGAN, we set $\lambda_{rls}$ to 0.4 and start-up at the beginning of training discriminators.\\
\indent We have used three objective metrics and one subjective metric to evaluate our methods. To measure the accuracy of the waveform that vocoder transforms, we use the MCD\cite{kubichek1993mel} and FFE\cite{chu2009reducing} between the ground truth and the synthesized waveform. For the evaluation of waveform quality, we measured PESQ\cite{rix2001perceptual}, and use MOS to evaluate subjective synthesis quality.
\subsection{Ablation Study}
We conducted an ablation study to analyze the effect of the proposed methods on this Chinese dataset. Starting from the baseline model, MelGAN and PWGAN, which apply LSGAN and STFT loss, we added each of the proposed methods one at a time measuring MCD, FFE, and PESQ. Table \ref{tab:results} displays the results. Additionally, we observed that using a large batch size greatly speeds up MelGAN training and increases the quality of synthesized waveforms.

\subsection{Comparison on MOS}
To compare subjective speech quality between LSGAN-based vocoders and PRLSGAN-based vocoders, we measured the MOS score of the speech waveforms synthesized by each vocoder. To perform a fair comparison, we randomly selected 10 utterances\footnote{The audio samples are presented in the following URL: https://anonymous1086.github.io/prlsgan-vocoder/} from a test set for MOS testing and 20 native Mandarin speakers participated in the listening test. The results of the subjective MOS evaluation are presented in Table \ref{tab:results}. The results show that the proposed PRLSGAN vocoders consistently outperformed the typical MelGAN and Parallel WaveGAN in MOS scoring, indicating the effectiveness of combining PRLSGAN with MelGAN and PWGAN frameworks. 

\section{Conclusions}
In this work, we have proposed PRLSGAN, an improved GAN framework that considers the pointwise relative gap between the truism score of the generator and discriminator. We design a novel pointwise adversarial loss to increase the difficulty of the min-max adversarial process, forcing the generator to refine on its local artifacts. Our experimental results have shown that PRLSGAN can be seamlessly adapted into MelGAN or Parallel WaveGAN based neural vocoders, achieving a great performance gain, while keeping their original inference speed.

\bibliographystyle{IEEEtran}

\bibliography{template.bbl}


\end{document}